\documentclass{article}
\usepackage{ijcai19}
\usepackage{times}
\usepackage{soul}
\usepackage{url}
\usepackage[utf8]{inputenc}
\usepackage[small]{caption}
\usepackage{graphicx}
\usepackage{amsmath}
\usepackage{amsfonts}
\usepackage{booktabs}
\usepackage{algorithm}
\usepackage{algorithmic}
\usepackage{subcaption}
\usepackage{multirow,epstopdf,array}

\usepackage{color}
\usepackage{soul, xcolor}

\title{Recent Advances in Diversified Recommendation}

\author{
Qiong Wu$^{1,2}$\and
Yong Liu$^{1,2}$\and
Chunyan Miao$^{1,2,3}$\and
Yin Zhao$^4$\and
Lu Guan$^4$\and
Haihong Tang$^4$\and
\\
\affiliations
$^1$Alibaba-NTU Singapore Joint Research Institute\\
$^2$The Joint NTU-UBC Research Centre of Excellence in Active Living for the Elderly (LILY)\\
$^3$School of Computer Science and Engineering, Nanyang Technological University\\
$^4$Alibaba Group\\
\emails
\{wu.qiong, stephenliu, ascymiao\}@ntu.edu.sg,
\{zhaoyin.zy, guanlu.gl\}@alibaba-inc.com, piaoxue@taobao.com
}

\begin{document}

\setstcolor{blue}

\maketitle

\begin{abstract}
With the rapid development of recommender systems, accuracy is no longer the only golden criterion for evaluating whether the recommendation results are satisfying or not. In recent years, diversity has gained tremendous attention in recommender systems research, which has been recognized to be an important factor for improving user satisfaction. On the one hand, diversified recommendation helps increase the chance of answering ephemeral user needs. On the other hand, diversifying recommendation results can help the business improve product visibility and explore potential user interests. In this paper, we are going to review the recent advances in diversified recommendation. Specifically, we first review the various definitions of diversity and generate a taxonomy to shed light on how diversity have been modeled or measured in recommender systems. After that, we summarize the major optimization approaches to diversified recommendation from a taxonomic view. Last but not the least, we project into the future and point out trending research directions on this topic.

\end{abstract}

\section{Introduction}
The need for diversification manifests in various information retrieval tasks. This problem was first brought by search result diversification in the late 1990s, as queries submitted to search engines are often short and ambiguous~\cite{carbonell1998mmr}. For instance, the keyword `java' may have different interpretations, e.g., programming language, coffee, and island. Diversifying the search results can largely improve the chance to hit the user's real intent. In the last decade, various approaches have been proposed to promote the diversity in search results~\cite{drosou2010search,welch2011search,zhu2014learning,liang2014personalized,santos2015search,xu2017directly,xia2017adapting}.

In early 2000s, diversity began to gain attentions from recommender system researchers, and it was recognized to be an important factor to improve user satisfaction for the recommendation results~\cite{bradley2001improving}. Differing from search engines, recommender systems do not have such privilege to peep into the user's current intent, since no keyword is provided. What recommender systems often utilize is the user's past interactions (e.g., views, clicks, and purchases) with items (e.g., books, movies, news, and products), from which to infer the user's general preferences and interests, and then select items that match most with (most relevant to) his/her preferences and interests. However, at the early stage, many of the diversification techniques in recommender systems were initially borrowed from search result diversification. Without loosing generality, in this paper, we root in recommender systems to review the recent development in diversification techniques.

Generally speaking, in recommender systems, there are two main lines of approaches, i.e, content-based approaches and collaborative filtering-based approaches. These approaches, in essence, are similarity-based, which may often yield sub-optimal results. For example, content-based approaches tend to produce items matching the user's interests but cover a very narrow scope of topics. Collaborative filtering based approaches often favor popular items, and some of the long-tail items never get the chance to be recommended~\cite{ashkan2015optimal}. Diversified recommendation is thus proposed to address the above mentioned problems. On the one hand, diversifying recommendation results may help to increase the chance of answering ephemeral user needs, particularly important for users with eclectic interests. On the other hand, diversified recommendations can help the business to improve product visibility by increasing the exposure rate of those long-tail items.

We are motivated by the above to review the recent advances in diversified recommendation. Particularly, we focus on providing meaningful taxonomies that could be useful for classifying, correlating, and clarifying existing approaches. Such taxonomies are still missing in all the recent surveys on similar topics~\cite{chakraborty2016survey,han2017survey,kunaver2017diversity}. Specifically, we first review various definitions of diversity and generate a taxonomy to give an overview of how diversity has been modeled or measured in recommender systems. Then, we summarize the popular optimization approaches in diversified recommendation from a taxonomic view. Last but not the least, based on the thorough review of existing methods, we project into the future and point out several trending research directions on this topic.

\section{Definition of Diversity}
Diversity in recommender systems can be viewed at either \emph{individual} or \emph{aggregate} level. Individual diversity refers to the diversity of recommendations for a given user, while aggregate diversity refers to the diversity of recommendations across all users. Although aggregate diversity is a combinatory view of individual diversity, a higher individual diversity does not necessarily imply a higher aggregate diversity~\cite{adomavicius2012improving}. For example, if the system recommends the same set of five distinct items to all users, the recommendation list for each user is very diverse (i.e., high individual diversity). However, the system can only recommend five items out of the entire item pool, thus, the aggregate diversity is low.

From another perspective, individual diversity focuses on the problem of how to maximize item novelty in face of already recommended ones when generating the recommendation list, while aggregate diversity can be viewed as a problem of how to improve the ability of a recommender system to recommend long-tail items. A few works have been proposed to improve the aggregate diversity of the recommendation results~\cite{brynjolfsson2010research,adomavicius2011maximizing,adomavicius2012improving,oestreicher2012recommendation}, countering the effect of item popularity.

\begin{figure}[t!]
   \centering
   \includegraphics[width=0.48\textwidth]{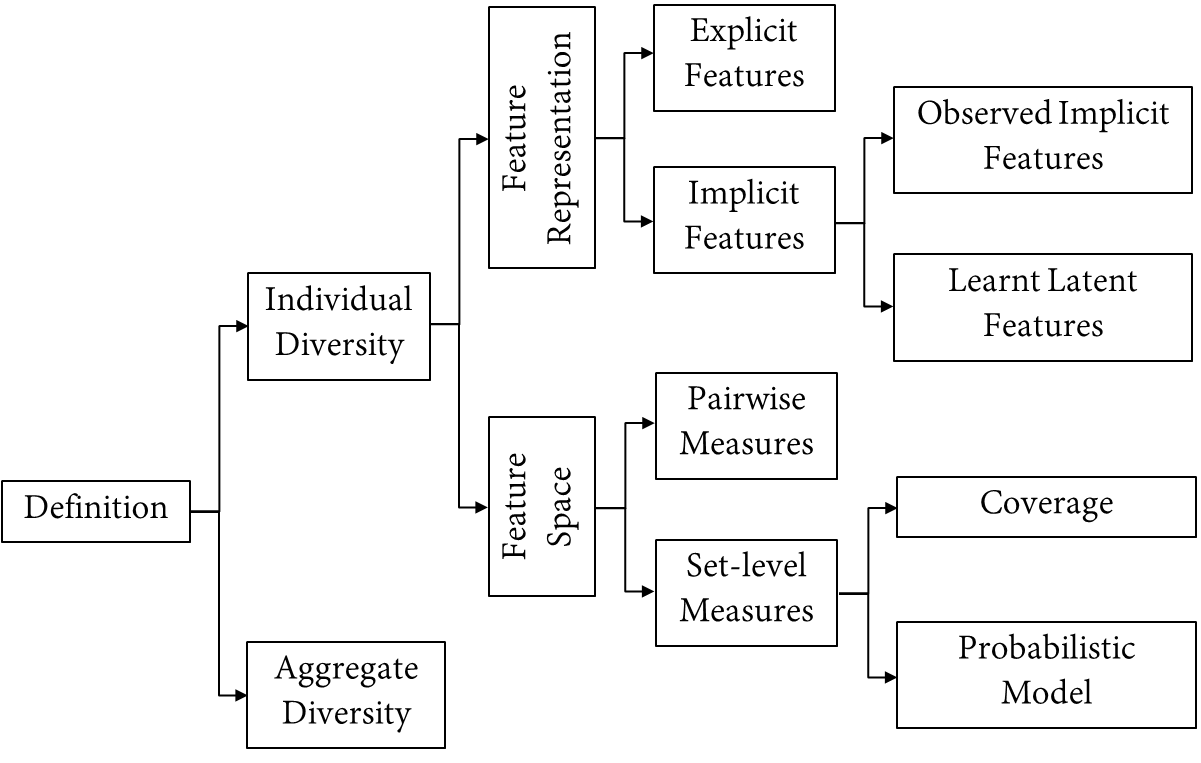}
   \caption{A taxonomy of definitions for diversity.}\label{fig:def}
\end{figure}

Simultaneously, a much larger body of research has been conducted to improve individual diversity, maximizing the between-item novelty. Therefore, in this paper, we focus on individual diversity, which we will simply refer as diversity throughout the paper, unless explicitly specified otherwise. In the rest of this section, we will review various lines of definitions for diversity in recommender systems and propose a taxonomy, as shown in Figure~\ref{fig:def}, which could encompass different views for diversity.

\subsection{Explicit v.s. Implicit Features}
The definition of diversity often involves a dissimilarity measure between items, and the first line of views split along whether explicit or implicit features are used for comparison. Specifically, \emph{Explicit features} refer to the attributes (e.g., genre, brands, and prices) or semantic taxonomy (e.g., topic hierarchy) describing items~\cite{ziegler2005improving}. Such information are often used by content-based filtering to perform similarity match between items. However, explicit features may not always be available, and in these cases, implicit features can be used to describe the properties of items. \emph{Implicit features} can be again categorized into two types: observed and learned. \emph{Observed implicit features} refer to some user generated data, for example, view, click, purchase records, etc. Such information have also been commonly used as the implicit feedback in recommender systems, as evidence for collaborative filtering~\cite{rendle2009bpr}. In addition, item features can also be learnt from the implicit or explicit feedbacks, where explicit feedbacks usually refer to the ratings in recommender systems, through latent factor models~\cite{koren2009matrix}. The latent factors comprise a computerized alternative to the aforementioned explicit features. The beauty of latent factors is that they can measure obvious dimensions as described by the explicit features, as well as less well-defined dimensions or completely uninterpretable dimensions that are not able to be captured by the explicit features~\cite{koren2009matrix}. These latent features are referred to as \emph{learnt latent features} in this work.

\subsection{Pairwise v.s. Set-level Measures} \label{dpp}
The second line of views split along whether diversity is defined based on a pairwise comparison between item features (no matter explicit or implicit), or based on a view of the entire set of recommended items as a whole.

\emph{Pairwise measures} usually define a dissimilarity function between items and use the average dissimilarity to characterize the diversity of a recommendation list~\cite{carbonell1998mmr,bradley2001improving,zhang2008avoiding}. The dissimilarity metric can be defined based on explicit features~\cite{ziegler2005improving,chandar2013preference,wu2016social}, or implicit features~\cite{qin2013promoting,cheng2017learning}.

\emph{Set-level measures} define the utility of the entire set of recommended items as a whole. One of the widely used set-level diversity metrics is \emph{coverage}. Coverage can be measured in proportionate to the number of distinct items~\cite{wu2016relevance,puthiya2016coverage} or topics~\cite{vargas2014coverage,ashkan2015optimal} covered by the recommended items. Another set-level diversity measure gaining rapid popularity in recent years is a \emph{probabilistic model} called Determinantal Point Process (DPP). DPP originated from quantum physics to give the distributions of fermion systems in thermal equilibrium. As DPP describes the repulsion of fermions precisely, it is natural for modeling diversity. DPP maintains a semi-definite kernel matrix $\mathbf{L}$ indexed by the entire candidate set of items, whose diagonal entries capture the quality of items while off-diagonal entries measure the similarity between items. Once $\mathbf{L}$ is determined, the probability of observing any subset of items $\mathcal{S}$ is proportionate to $\det(\mathbf{L}_{\mathcal{S}})$. Due to the property of matrix determinant, more diverse items get higher probability to be sampled together. The DPP kernel matrix $\mathbf{L}$ can be viewed as a model of the entire feature space of all items, which can be learnt from data~\cite{gillenwater2014expectation,mariet2015fixed,gartrell2016bayesian,gartrell2017low,wilhelm2018practical} or carefully constructed from explicit or implicit item features~\cite{chen2018fast}.

\section{Optimization of Diversity}
Diversified recommendation is generally treated as a bi-criterion optimization problem, in which one seeks to maximize the overall relevance of a recommendation list, while minimizing the redundancy between the recommended items~\cite{gollapudi2009axiomatic}. Therefore, one commonly shared objective among most diversified recommendation methods is to maintain a proper trade-off between the relevance and the diversity of the recommendation results~\cite{carbonell1998mmr}.

\begin{figure}[t!]
   \centering
   \includegraphics[width=0.48\textwidth]{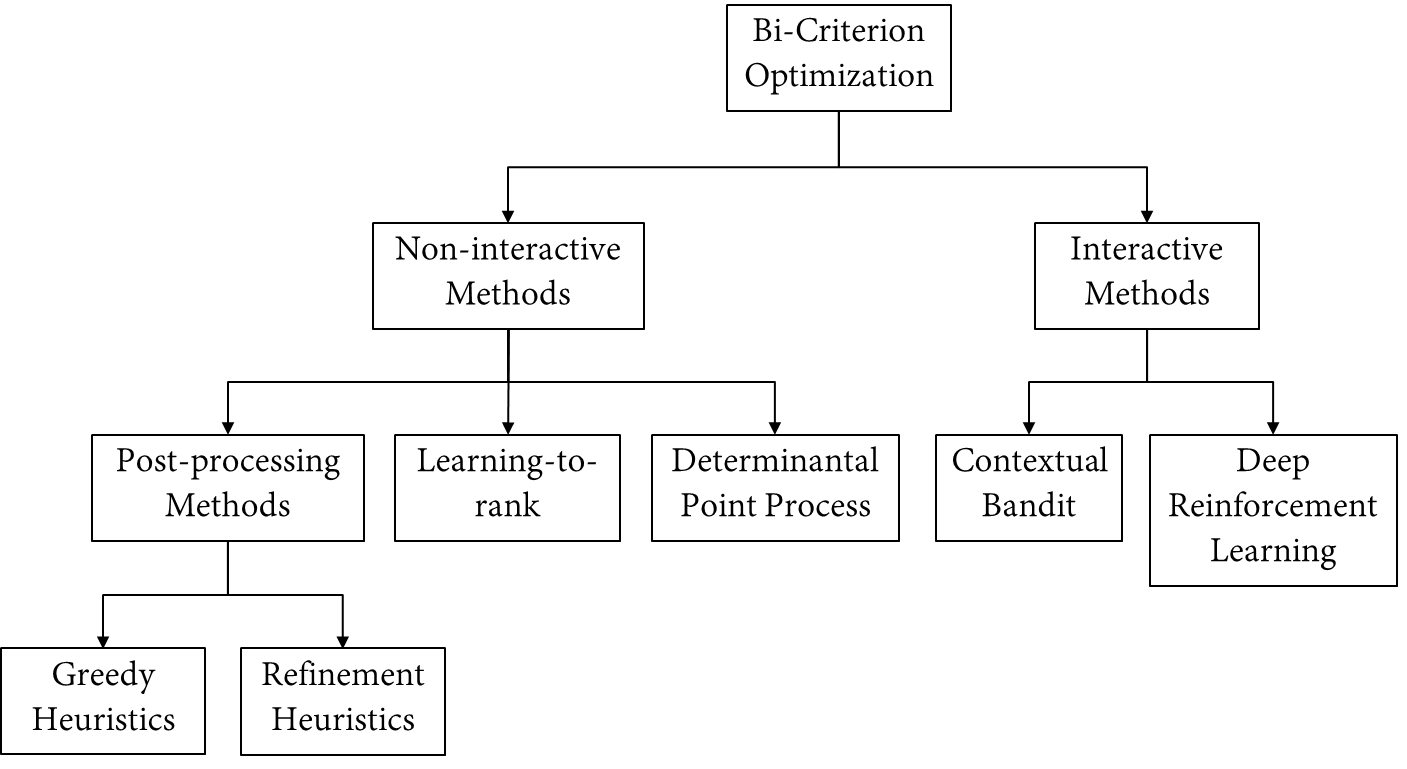}
   \caption{A Taxonomy of major optimization methods for diversity.}\label{fig:opt}
\end{figure}

Various approaches have been proposed to realize the above mentioned objective in recommender systems. In general, we can categorize the major approaches into two broad classes, based on whether the optimization is done in an offline manner or an online manner. In the offline setting, recommender systems usually generate recommendations based on the historical interactions between users and items. With such data, there are three major approaches for trading-off between relevance and diversity, i.e., post-processing methods, learning-to-rank methods, and determinantal point process based methods. These offline methods, however, usually ignore the interactions between users and the recommender system, and thus cannot take the users' immediate and long-term feedback to improve the recommender system's performance. On the other hand, in the online setting, a recommender system explicitly models the interactions between users and the recommender system, which allows the recommender system to update its recommendation policy based on users' feedback in an interactive manner. Two popular approaches have been used to diversify recommendation results for online interactive recommendations, i.e., contextual bandit and deep reinforcement learning. We summarize this taxonomy in Figure~\ref{fig:opt}. Next, we will discuss the major methods following this taxonomy.

\subsection{Non-interactive Methods}
Non-interactive methods usually train an offline model using historical user-item interaction data and then generate recommendations based on the trained model. In this section, we review three lines of non-interactive methods for diversified recommendation, i.e., post-processing methods, learning-to-rank, and determinantal point process.
\subsubsection{Post-processing Methods}
The earliest works on diversified recommendations usually take a heuristic post-processing approach to strive a balance between relevance and diversity in the top-N recommendation list~\cite{qin2013promoting}. These approaches can be classified into two classes: 1) greedy heuristics, where the recommendation list is greedily constructed one-by-one by maximizing a marginal relevance; and 2) refinement heuristics, where items are first ranked based on a relevance metric and then refined by introducing a diversity metric.

The pioneering approach on greedy heuristics is Maximal Marginal Relevance (MMR)~\cite{carbonell1998mmr}, which first represents relevance and diversity by independent metrics and then uses the notion of marginal relevance to combine the two metrics with a trade-off parameter. MMR creates a diversified ranking of items by choosing an item in each interaction such that it maximizes the marginal relevance. Other greedy heuristics methods vary in the definition of the marginal relevance, often in the form of a submodular objective function, which can be solved greedily with an approximation to the optimal solution. For example, in~\cite{qin2013promoting}, an entropy regularizer is proposed to capture the notion of diversity which satisfies monotonicity and submodularity. It is then combined with the modular rating set function, which gives a submodular objective function. This submodular objective is maximized approximately by greedy algorithm. In~\cite{ashkan2015optimal}, another form of submodular objective function called diversity-weighted utility maximization was proposed, and it was then maximized by a greedy algorithm. In~\cite{sha2016framework}, a submodular objective function was proposed to combine relevance, coverage of user's interests, and the diversity between items.

Refinement heuristics usually re-rank a pre-ranked item list through some post-processing actions. For example,~\cite{ziegler2005improving} defines a similarity metric based on the taxonomy information to compute an intra-list similarity for determining the overall diversity of the recommendation list, and increases diversity by merging a dissimilarity rank with the original rank.~\cite{zhang2008avoiding} defines item similarity based on the feature space (explicit or implicit features) and optimizes diversity by relaxing the bi-criterion optimization problem to a trust-region problem.~\cite{yu2009takes} proposes the notion of explanation-based diversity and then develops two re-ranking strategies, i.e., swap and greedy, to promote diversity in top-N recommendations.~\cite{boim2011diversification} adopts item-based CF for recommendation and represents each item with a vector of ratings from users. Pearson's correlation coefficient is used to measure the similarity between items and the diversity of the recommendation list is improved through the construction of a priority cover-tree.

\subsubsection{Learning-to-rank}
Post-processing methods mainly play some tricks at the ranking stage. However, such a strategy is suboptimal as cumulative loss on ranking performance may be generated due to error on each step, and extensive tuning of the trade-off parameter may be required~\cite{li2017learning}. To mitigate these problems, another group of research proposes to directly learn an optimal ranking for each user, which is called Learning-To-Rank (LTR).

The LTR methods target at the right recommendation order to each user rather than precise score prediction on each candidate item. For diversified recommendation, the main problem of LTR becomes defining an ideal ranking that considers both relevance and diversity as the ground-truth for learning. Several recent works have been done in this direction, for example, ~\cite{cheng2017learning} treats each user as a training instance, which is labeled by a set of relevant and diverse items empirically determined. An automatic heuristic labeling method is proposed to trade-off between relevance and diversity. LTR is then becoming a supervised learning problem.~\cite{li2017learning} defines an ideal recommendation list by proposing a score formulation to model contribution of a candidate item at a particular recommendation position, which unifies both relevance and diversity. LTR becomes a problem of optimizing the score model with a pairwise loss function.

\subsubsection{Determinantal Point Process}
Most of post-processing and LTR methods adopt pairwise measures of diversity, which is suboptimal as they ignore the correlations between items~\cite{chen2018fast}. Moreover, with the current techniques, similar items may have similar quality scores, which make it expensive to score every possible permutation of the ranked list to generate the ideal recommendation that balances relevance and diversity~\cite{wilhelm2018practical}. Another line of research strives to address the above mentioned issues through a probability model of diversity, called Determinantal Point Process (DPP).

As introduced in the previous section~\ref{dpp}, DPP maintains a kernel matrix $\mathbf{L}$ indexed by the entire candidate set of items, whose diagonal entries capture the quality of items while off-diagonal entries measure the similarity between items. The $\mathbf{L}$ kernel matrix models the entire feature space of items and captures the correlations between them, which elegantly solves the first problem. Once $\mathbf{L}$ is determined, various efficient sequential sampling techniques can be applied to generate a list of diverse and relevant items for recommendation~\cite{chen2018fast,wilhelm2018practical}, which smartly solves the second problem.

There are two approaches for determining $\mathbf{L}$, either learnt from data or constructed empirically. For example,~\cite{wilhelm2018practical} proposes to learn $\mathbf{L}$ from observed sets of historically interacted items, and then uses an approximate inference algorithm for efficiently sampling from the learnt $\mathbf{L}$. Several other works also adopt similar ideas~\cite{gillenwater2014expectation,mariet2015fixed,gartrell2016bayesian,gartrell2017low}, but mainly use DPP for basket completion tasks rather than diversifying recommendation.~\cite{chen2018fast} proposes to construct $\mathbf{L}$ using quality scores and item features. A trade-off parameter is introduced to balance the consideration of relevance and diversity when constructing $\mathbf{L}$. Moreover, a fast greedy Maximum A Posteriori (MAP) inference is developed in this work to efficiently sample from constructed $\mathbf{L}$.

\subsection{Interactive Methods}
Interactive methods usually train an online recommendation model by repeating the following steps: push the recommendation result, collect feedback, and update model. In this section, we review two lines of interactive methods for diversified recommendation, i.e., contextual bandit and deep reinforcement learning.
\subsubsection{Contextual Bandit}
The non-interactive methods make use of historical interaction data between users and items to infer users' general preferences for generating relevant recommendations. However, in real world, new users and new items frequently join the system over time, and the profiles of users update dynamically~\cite{wang2017biucb}, which is called the cold-start problem in recommender systems. Such a problem can be naturally modeled as a Contextual Multi-Armed Bandit (CMAB) problem. Taking movie recommendation as an example, in CMAB setting, given a new user, the recommender system can repeatedly provide the user with a set of movies (called a super arm) and collect his/her feedback in multiple rounds. The CMAB algorithm helps to decide whether to try some new movies (i.e., exploration) or stick on the movies that the user prefers (i.e., exploitation) in the next round based on the user's feedback.

In CMAB setting, diversity is often promoted by designing a relevance-diversity bi-criterion reward function for the super arm. For example, in~\cite{qin2014contextual}, entropy regularizer is incorporated into the reward function (originally consisting of relevance alone) to promote diversity of the selected super arm. In~\cite{wang2017biucb}, three types of diversified reward function have been explored in a CMAB setting, including MMR~\cite{carbonell1998mmr}, entropy regularizer~\cite{qin2014contextual} and temporal user-based switching~\cite{lathia2010temporal}.

\subsubsection{Deep Reinforcement Learning}
Contextual bandit methods generally assume that the state of the user is invariant. However, in real world, the user's preference is dynamically changing over time. This information can be captured naturally in a Reinforcement Learning (RL) framework. In RL setting, a recommender system can be viewed as an agent. It senses the state of the environment (i.e., the user's current preferences), and generate an action (i.e., recommend a set of items) accordingly. It then collects rewards based on the feedbacks provided by the user and updates its recommendation policy by optimizing the current rewards or/and the cumulative future rewards.

In RL setting, diversity has been promoted by employing efficient exploration-exploitation strategies. For example, in~\cite{zheng2018drn}, a dueling bandit gradient descent algorithm is adopted to do exploration in a deep reinforcement learning framework, which chooses random item candidates in the neighborhood of the current recommender. This exploration strategy can promote diversity while avoiding recommending totally unrelated items and thus preserving recommendation accuracy.

\vspace{0.5em}
\noindent{\bf Summary.} In short summary, LTR and DPP are two recent trending approaches to solve the diversified recommendation problem in the offline setting. Overall speaking, approaches for diversifying recommendation in the online setting is less well studied than that in the offline setting. Deep RL is a trending framework to study diversified interactive recommendation in the online setting.

\section{Future Directions}
Although various approaches have been explored for diversified recommendation, there are still many open questions and challenges to be studied. In this section, we point out five important future research directions on this topic.

\subsection{Personalized Diversity}
Since diversified recommendation is often treated as a bi-criterion
optimization problem, the objective function (or the reward function in interactive recommendations) usually involves a trade-off parameter between diversity and accuracy. In majority of the diversified recommendation approaches, this parameter is global in nature, which means the same parameter is applied for all users to balance diversity and accuracy~\cite{han2017survey}. Defining such a trade-off parameter in the objective function can be avoided in the learning-to-rank methods, as the optimal ranking is directly learnt from ground-truth sets. However, the same problem pop up when determining a proper ground-truth set that balancing diversity and accuracy.

It is natural that different users may have different preferences to diversity. Some of the users may have very focused interests while others may have a very broad scope of interests. Hence, treating the trade-off parameter in a global manner can be sub-optimal. For example, blindly pursing high diversity for focused users may definitely hurt the accuracy of recommendation results. However, it is still an open question on how to learn users' personal preferences for the diversity of recommendation results and to personalize the trade-off parameters for each individual user.

\subsection{Temporal Diversity}
Most of the current diversified recommendation methods focus on improving the diversity of recommendation results in one recommendation session, but ignoring the diversity of recommendation results across multiple recommendation sessions. The latter problem was first brought up in~\cite{lathia2010temporal}, and was defined as a problem of temporal diversity. Temporal diversity address the problem of the same item being recommended to the user over and over again, or whether the recommended items present a certain level of novelty~\cite{chakraborty2016survey}.

The problem of temporal diversity is especially prominent in non-interactive recommender systems, as the same set of items will be repeatedly recommended to the user unless the recommendation model updates based on newly generated user-item interactions. In~\cite{lathia2010temporal}, various algorithm switching or re-ranking strategies have been explored to promote temporal diversity in a number of non-interactive recommendation models (e.g., CF, k-NN, SVD). However, temporal diversity can be more naturally modeled in the interactive recommendation models, as interactive recommendation models are usually built on a session basis. Moreover, in interactive recommendation models, temporal diversity can be transformed into an exploration-exploitation problem, as more exploration may induce a higher level of temporal diversity. Yet, few studies have been conducted in interactive recommendation models to promote temporal diversity.

\subsection{Explainable Diversity}
One of the common problem faced by traditional recommender systems is that they only provide a set of items for recommendation, but do not provide explanations to make the user or system designer aware of why such items are recommended. However, meaningful and fashionable explanations can help to improve the effectiveness, efficiency and persuasiveness of the recommendation results. In recent years, explainable recommendation is gaining rapid popularity to enhance the transparency of the logic behind sophisticated recommendation models~\cite{zhang2018explainable}.

Most of the current explainable recommendation approaches focus on finding associations between the recommendation results and relevant users or items, to generate various forms of explanations in textual or visual formats~\cite{zhang2018explainable}. However, it is still a void field to study how diversified recommendations can be properly explained. This may involve investigating new explanation strategies or rationales, as diversity not only highlights the relevance of individual items, but also the novelty of items in face of already recommended ones.

\subsection{Visual Diversity}
In many real world applications of recommender systems, e.g., e-commerce websites, the recommendation results are often shown as a page of item displays~\cite{zhao2018deep}. The users can quickly scan a large number of potential purchases through visual browsing. Hence, a carefully designed item display strategy can potentially enhance users' online visual browsing experience and help them in item discovery~\cite{teo2016adaptive}.

Towards this end,~\cite{teo2016adaptive} has made the first attempt to diversify visual shopping experience which adopts a submodular diversification framework to re-rank the top scoring items based on category information. However, this approach is still item re-ranking by nature, which does not consider other important information that may influence users' visual experiences, such as the relative positions or graphics of displayed items. Hence, visual diversity remains another interesting domain to be investigated in future research.

\subsection{Psychology-driven Diversity}
If we look at the major body of research in recommender systems, they are data-driven in essence. However, data are produced by users, which only capture the decision of users but never the decision-making process of them. So the question is, can we make better predictive models by understanding the users better? This may need the synergy of theories from multiple disciplines, such as psychology, social science, neuroscience, and of course, data science.

Human beings naturally prefer novelty, as driven by their intrinsic curiosity~\cite{wu2013curiosity}, which leads to a need for diverse stimuli.~\cite{wu2016social} has made the first attempt to generate diversified recommendations by modeling users' social curiosity, borrowing ideas from human psychology on how our curiosity is stimulated through surprises. Social curiosity not only helps to generate more diversified recommendation lists, but also helps to reveal many long-tailed items that have been surprisingly rated high by friends. However, such ideas are still in its infancy in recommender systems research and open up new possible ways for improving the recommendation performance and user satisfaction.

\section{Conclusion}
Diversity plays an important role in making the recommendation results more interesting and meaningful for the benefit of both end users and business providers. It has been widely studied in the past two decades by recommender system researchers. Yet, there are still many open problems remaining to be addressed in this domain of research. In this paper, we have reviewed and analyzed the recent advances in diversified recommendation in a taxonomic way and shed light on new research perspectives. More specifically, we first reviewed the various lines of definitions for diversity in recommender systems and created a comprehensive taxonomy to give an overview on how diversity have been modeled or measured. We then summarized the major optimization approaches for diversified recommendation and created a taxonomy for classifying, correlating, and clarifying existing approaches. Based on the thorough review, we projected into the future and highlighted five important future research directions that remain open for discussion, in terms of personalization, temporal characteristics, explainability, visual experiences, and human psychology, respectively.

\section*{Acknowledgments}
This research is partially supported by the Alibaba-NTU Singapore Joint Research Institute.
\bibliographystyle{named}
\bibliography{ijcai19}

\begin{thebibliography}{}

\bibitem[\protect\citeauthoryear{Adomavicius and
  Kwon}{2011}]{adomavicius2011maximizing}
Gediminas Adomavicius and YoungOk Kwon.
\newblock Maximizing aggregate recommendation diversity: A graph-theoretic
  approach.
\newblock In {\em DiveRS}, pages 3--10, 2011.

\bibitem[\protect\citeauthoryear{Adomavicius and
  Kwon}{2012}]{adomavicius2012improving}
Gediminas Adomavicius and YoungOk Kwon.
\newblock Improving aggregate recommendation diversity using ranking-based
  techniques.
\newblock {\em IEEE Transactions on Knowledge and Data Engineering},
  24(5):896--911, 2012.

\bibitem[\protect\citeauthoryear{Ashkan \bgroup \em et al.\egroup
  }{2015}]{ashkan2015optimal}
Azin Ashkan, Branislav Kveton, Shlomo Berkovsky, and Zheng Wen.
\newblock Optimal greedy diversity for recommendation.
\newblock In {\em IJCAI}, pages 1742--1748, 2015.

\bibitem[\protect\citeauthoryear{Boim \bgroup \em et al.\egroup
  }{2011}]{boim2011diversification}
Rubi Boim, Tova Milo, and Slava Novgorodov.
\newblock Diversification and refinement in collaborative filtering
  recommender.
\newblock In {\em CIKM}, pages 739--744. ACM, 2011.

\bibitem[\protect\citeauthoryear{Bradley and
  Smyth}{2001}]{bradley2001improving}
Keith Bradley and Barry Smyth.
\newblock Improving recommendation diversity.
\newblock In {\em AICS}, pages 85--94, 2001.

\bibitem[\protect\citeauthoryear{Brynjolfsson \bgroup \em et al.\egroup
  }{2010}]{brynjolfsson2010research}
Erik Brynjolfsson, Yu~Hu, and Michael~D Smith.
\newblock Research commentary—long tails vs. superstars: The effect of
  information technology on product variety and sales concentration patterns.
\newblock {\em Information Systems Research}, 21(4):736--747, 2010.

\bibitem[\protect\citeauthoryear{Carbonell and
  Goldstein}{1998}]{carbonell1998mmr}
Jaime Carbonell and Jade Goldstein.
\newblock The use of mmr, diversity-based reranking for reordering documents
  and producing summaries.
\newblock In {\em SIGIR}, pages 335--336, 1998.

\bibitem[\protect\citeauthoryear{Chakraborty and
  Verma}{2016}]{chakraborty2016survey}
Jayeeta Chakraborty and Vijay Verma.
\newblock A survey of diversification techniques in recommendation systems.
\newblock In {\em DMAC}, pages 35--40, 2016.

\bibitem[\protect\citeauthoryear{Chandar and
  Carterette}{2013}]{chandar2013preference}
Praveen Chandar and Ben Carterette.
\newblock Preference based evaluation measures for novelty and diversity.
\newblock In {\em SIGIR}, pages 413--422, 2013.

\bibitem[\protect\citeauthoryear{Chen \bgroup \em et al.\egroup
  }{2018}]{chen2018fast}
Laming Chen, Guoxin Zhang, and Eric Zhou.
\newblock Fast greedy map inference for determinantal point process to improve
  recommendation diversity.
\newblock In {\em NIPS}, pages 5623--5634, 2018.

\bibitem[\protect\citeauthoryear{Cheng \bgroup \em et al.\egroup
  }{2017}]{cheng2017learning}
Peizhe Cheng, Shuaiqiang Wang, Jun Ma, Jiankai Sun, and Hui Xiong.
\newblock Learning to recommend accurate and diverse items.
\newblock In {\em WWW}, pages 183--192, 2017.

\bibitem[\protect\citeauthoryear{Drosou and Pitoura}{2010}]{drosou2010search}
Marina Drosou and Evaggelia Pitoura.
\newblock Search result diversification.
\newblock {\em SIGMOD record}, 39(1):41--47, 2010.

\bibitem[\protect\citeauthoryear{Gartrell \bgroup \em et al.\egroup
  }{2016}]{gartrell2016bayesian}
Mike Gartrell, Ulrich Paquet, and Noam Koenigstein.
\newblock Bayesian low-rank determinantal point processes.
\newblock In {\em RecSys}, pages 349--356, 2016.

\bibitem[\protect\citeauthoryear{Gartrell \bgroup \em et al.\egroup
  }{2017}]{gartrell2017low}
Mike Gartrell, Ulrich Paquet, and Noam Koenigstein.
\newblock Low-rank factorization of determinantal point processes for
  recommendation.
\newblock In {\em AAAI}, pages 1912--1918, 2017.

\bibitem[\protect\citeauthoryear{Gillenwater \bgroup \em et al.\egroup
  }{2014}]{gillenwater2014expectation}
Jennifer~A Gillenwater, Alex Kulesza, Emily Fox, and Ben Taskar.
\newblock Expectation-maximization for learning determinantal point processes.
\newblock In {\em NIPS}, pages 3149--3157, 2014.

\bibitem[\protect\citeauthoryear{Gollapudi and
  Sharma}{2009}]{gollapudi2009axiomatic}
Sreenivas Gollapudi and Aneesh Sharma.
\newblock An axiomatic approach for result diversification.
\newblock In {\em WWW}, pages 381--390. ACM, 2009.

\bibitem[\protect\citeauthoryear{Han and Yamana}{2017}]{han2017survey}
Jungkyu Han and Hayato Yamana.
\newblock A survey on recommendation methods beyond accuracy.
\newblock {\em IEICE Transactions on Information and Systems},
  100(12):2931--2944, 2017.

\bibitem[\protect\citeauthoryear{Koren \bgroup \em et al.\egroup
  }{2009}]{koren2009matrix}
Yehuda Koren, Robert Bell, and Chris Volinsky.
\newblock Matrix factorization techniques for recommender systems.
\newblock {\em Computer}, (8):30--37, 2009.

\bibitem[\protect\citeauthoryear{Kunaver and
  Po{\v{z}}rl}{2017}]{kunaver2017diversity}
Matev{\v{z}} Kunaver and Toma{\v{z}} Po{\v{z}}rl.
\newblock Diversity in recommender systems--a survey.
\newblock {\em Knowledge-Based Systems}, 123:154--162, 2017.

\bibitem[\protect\citeauthoryear{Lathia \bgroup \em et al.\egroup
  }{2010}]{lathia2010temporal}
Neal Lathia, Stephen Hailes, Licia Capra, and Xavier Amatriain.
\newblock Temporal diversity in recommender systems.
\newblock In {\em SIGIR}, pages 210--217. ACM, 2010.

\bibitem[\protect\citeauthoryear{Li \bgroup \em et al.\egroup
  }{2017}]{li2017learning}
Shuang Li, Yuezhi Zhou, Di~Zhang, Yaoxue Zhang, and Xiang Lan.
\newblock Learning to diversify recommendations based on matrix factorization.
\newblock In {\em DASC/PiCom/DataCom/CyberSciTech}, pages 68--74. IEEE, 2017.

\bibitem[\protect\citeauthoryear{Liang \bgroup \em et al.\egroup
  }{2014}]{liang2014personalized}
Shangsong Liang, Zhaochun Ren, and Maarten De~Rijke.
\newblock Personalized search result diversification via structured learning.
\newblock In {\em SIGKDD}, pages 751--760. ACM, 2014.

\bibitem[\protect\citeauthoryear{Mariet and Sra}{2015}]{mariet2015fixed}
Zelda Mariet and Suvrit Sra.
\newblock Fixed-point algorithms for learning determinantal point processes.
\newblock In {\em ICML}, pages 2389--2397, 2015.

\bibitem[\protect\citeauthoryear{Oestreicher-Singer and
  Sundararajan}{2012}]{oestreicher2012recommendation}
Gal Oestreicher-Singer and Arun Sundararajan.
\newblock Recommendation networks and the long tail of electronic commerce.
\newblock {\em Mis quarterly}, pages 65--83, 2012.

\bibitem[\protect\citeauthoryear{Puthiya~Parambath \bgroup \em et al.\egroup
  }{2016}]{puthiya2016coverage}
Shameem~A Puthiya~Parambath, Nicolas Usunier, and Yves Grandvalet.
\newblock A coverage-based approach to recommendation diversity on similarity
  graph.
\newblock In {\em RecSys}, pages 15--22, 2016.

\bibitem[\protect\citeauthoryear{Qin and Zhu}{2013}]{qin2013promoting}
Lijing Qin and Xiaoyan Zhu.
\newblock Promoting diversity in recommendation by entropy regularizer.
\newblock In {\em IJCAI}, pages 2698--2704, 2013.

\bibitem[\protect\citeauthoryear{Qin \bgroup \em et al.\egroup
  }{2014}]{qin2014contextual}
Lijing Qin, Shouyuan Chen, and Xiaoyan Zhu.
\newblock Contextual combinatorial bandit and its application on diversified
  online recommendation.
\newblock In {\em ICDM}, pages 461--469, 2014.

\bibitem[\protect\citeauthoryear{Rendle \bgroup \em et al.\egroup
  }{2009}]{rendle2009bpr}
Steffen Rendle, Christoph Freudenthaler, Zeno Gantner, and Lars Schmidt-Thieme.
\newblock Bpr: Bayesian personalized ranking from implicit feedback.
\newblock In {\em UAI}, pages 452--461, 2009.

\bibitem[\protect\citeauthoryear{Santos \bgroup \em et al.\egroup
  }{2015}]{santos2015search}
Rodrygo~LT Santos, Craig Macdonald, Iadh Ounis, et~al.
\newblock Search result diversification.
\newblock {\em Foundations and Trends{\textregistered} in Information
  Retrieval}, 9(1):1--90, 2015.

\bibitem[\protect\citeauthoryear{Sha \bgroup \em et al.\egroup
  }{2016}]{sha2016framework}
Chaofeng Sha, Xiaowei Wu, and Junyu Niu.
\newblock A framework for recommending relevant and diverse items.
\newblock In {\em IJCAI}, pages 3868--3874, 2016.

\bibitem[\protect\citeauthoryear{Teo \bgroup \em et al.\egroup
  }{2016}]{teo2016adaptive}
Choon~Hui Teo, Houssam Nassif, Daniel Hill, Sriram Srinivasan, Mitchell
  Goodman, Vijai Mohan, and SVN Vishwanathan.
\newblock Adaptive, personalized diversity for visual discovery.
\newblock In {\em RecSys}, pages 35--38, 2016.

\bibitem[\protect\citeauthoryear{Vargas \bgroup \em et al.\egroup
  }{2014}]{vargas2014coverage}
Sa{\'u}l Vargas, Linas Baltrunas, Alexandros Karatzoglou, and Pablo Castells.
\newblock Coverage, redundancy and size-awareness in genre diversity for
  recommender systems.
\newblock In {\em RecSys}, pages 209--216, 2014.

\bibitem[\protect\citeauthoryear{Wang \bgroup \em et al.\egroup
  }{2017}]{wang2017biucb}
Lu~Wang, Chengyu Wang, Keqiang Wang, and Xiaofeng He.
\newblock Biucb: A contextual bandit algorithm for cold-start and diversified
  recommendation.
\newblock In {\em ICBK}, pages 248--253. IEEE, 2017.

\bibitem[\protect\citeauthoryear{Welch \bgroup \em et al.\egroup
  }{2011}]{welch2011search}
Michael~J Welch, Junghoo Cho, and Christopher Olston.
\newblock Search result diversity for informational queries.
\newblock In {\em WWW}, pages 237--246. ACM, 2011.

\bibitem[\protect\citeauthoryear{Wilhelm \bgroup \em et al.\egroup
  }{2018}]{wilhelm2018practical}
Mark Wilhelm, Ajith Ramanathan, Alexander Bonomo, Sagar Jain, Ed~H Chi, and
  Jennifer Gillenwater.
\newblock Practical diversified recommendations on youtube with determinantal
  point processes.
\newblock In {\em CIKM}, pages 2165--2173, 2018.

\bibitem[\protect\citeauthoryear{Wu and Miao}{2013}]{wu2013curiosity}
Qiong Wu and Chunyan Miao.
\newblock Curiosity: From psychology to computation.
\newblock {\em ACM Computing Surveys (CSUR)}, 46(2):18, 2013.

\bibitem[\protect\citeauthoryear{Wu \bgroup \em et al.\egroup
  }{2016a}]{wu2016relevance}
Le~Wu, Qi~Liu, Enhong Chen, Nicholas~Jing Yuan, Guangming Guo, and Xing Xie.
\newblock Relevance meets coverage: A unified framework to generate diversified
  recommendations.
\newblock {\em ACM Transactions on Intelligent Systems and Technology (TIST)},
  7(3):39, 2016.

\bibitem[\protect\citeauthoryear{Wu \bgroup \em et al.\egroup
  }{2016b}]{wu2016social}
Qiong Wu, Siyuan Liu, Chunyan Miao, Yuan Liu, and Cyril Leung.
\newblock A social curiosity inspired recommendation model to improve
  precision, coverage and diversity.
\newblock In {\em WI}, pages 240--247, 2016.

\bibitem[\protect\citeauthoryear{Xia \bgroup \em et al.\egroup
  }{2017}]{xia2017adapting}
Long Xia, Jun Xu, Yanyan Lan, Jiafeng Guo, Wei Zeng, and Xueqi Cheng.
\newblock Adapting markov decision process for search result diversification.
\newblock In {\em SIGIR}, pages 535--544. ACM, 2017.

\bibitem[\protect\citeauthoryear{Xu \bgroup \em et al.\egroup
  }{2017}]{xu2017directly}
Jun Xu, Long Xia, Yanyan Lan, Jiafeng Guo, and Xueqi Cheng.
\newblock Directly optimize diversity evaluation measures: a new approach to
  search result diversification.
\newblock {\em ACM Transactions on Intelligent Systems and Technology (TIST)},
  8(3):41, 2017.

\bibitem[\protect\citeauthoryear{Yu \bgroup \em et al.\egroup
  }{2009}]{yu2009takes}
Cong Yu, Laks Lakshmanan, and Sihem Amer-Yahia.
\newblock It takes variety to make a world: diversification in recommender
  systems.
\newblock In {\em EDBT}, pages 368--378, 2009.

\bibitem[\protect\citeauthoryear{Zhang and Chen}{2018}]{zhang2018explainable}
Yongfeng Zhang and Xu~Chen.
\newblock Explainable recommendation: A survey and new perspectives.
\newblock {\em arXiv preprint arXiv:1804.11192}, 2018.

\bibitem[\protect\citeauthoryear{Zhang and Hurley}{2008}]{zhang2008avoiding}
Mi~Zhang and Neil Hurley.
\newblock Avoiding monotony: improving the diversity of recommendation lists.
\newblock In {\em RecSys}, pages 123--130, 2008.

\bibitem[\protect\citeauthoryear{Zhao \bgroup \em et al.\egroup
  }{2018}]{zhao2018deep}
Xiangyu Zhao, Long Xia, Liang Zhang, Zhuoye Ding, Dawei Yin, and Jiliang Tang.
\newblock Deep reinforcement learning for page-wise recommendations.
\newblock In {\em RecSys}, pages 95--103, 2018.

\bibitem[\protect\citeauthoryear{Zheng \bgroup \em et al.\egroup
  }{2018}]{zheng2018drn}
Guanjie Zheng, Fuzheng Zhang, Zihan Zheng, Yang Xiang, Nicholas~Jing Yuan, Xing
  Xie, and Zhenhui Li.
\newblock Drn: A deep reinforcement learning framework for news recommendation.
\newblock In {\em WWW}, pages 167--176, 2018.

\bibitem[\protect\citeauthoryear{Zhu \bgroup \em et al.\egroup
  }{2014}]{zhu2014learning}
Yadong Zhu, Yanyan Lan, Jiafeng Guo, Xueqi Cheng, and Shuzi Niu.
\newblock Learning for search result diversification.
\newblock In {\em SIGIR}, pages 293--302. ACM, 2014.

\bibitem[\protect\citeauthoryear{Ziegler \bgroup \em et al.\egroup
  }{2005}]{ziegler2005improving}
Cai-Nicolas Ziegler, Sean~M McNee, Joseph~A Konstan, and Georg Lausen.
\newblock Improving recommendation lists through topic diversification.
\newblock In {\em WWW}, pages 22--32, 2005.

\end{thebibliography}
\end{document}